\documentclass[prx,nofootinbib]{revtex4}

\usepackage{amsmath,srcltx}
\usepackage{amssymb}  
\usepackage{color}  
\usepackage{graphicx}
\usepackage{feynmp}
\usepackage{float}
\usepackage{mathrsfs} 
\begin{document}

\title{Quantum Mechanics in Wavelet Basis}

\author{Pavan Chawhan} \email{f20160431g@alumni.bits-pilani.ac.in} 

\author{Raghunath Ratabole} \email{ratabole@goa.bits-pilani.ac.in}

\affiliation{Department of Physics, Birla Institute of Technology and Science Pilani, K K Birla Goa Campus, Zuarinagar 403726, Goa, India}

%===================================================================
\begin{abstract}

We describe a multi-scale resolution approach to analyzing problems in Quantum Mechanics using Daubechies wavelet basis. The expansion of the wavefunction of the quantum system in this basis allows a natural interpretation of each basis function as a quantum fluctuation of a specific resolution at a particular location. The Hamiltonian matrix constructed in this basis describes couplings between different length scales and thus allows for intuitive volume and resolution truncation. In quantum mechanical problems with a natural length scale, one can get approximate solution of the problem through simple matrix diagonalization. We illustrate this approach using the example of the standard quantum mechanical simple harmonic oscillator. 
 
\end{abstract}
\maketitle
%===================================================================

%%%%%%%%%%%%%%%%%%%%%%%%%%%%%%%%%%%%%%%%%%%%%%
\section{Introduction}
%%%%%%%%%%%%%%%%%%%%%%%%%%%%%%%%%%%%%%%%%%%%%%
The solution of the Hamiltonian energy eigenvalue problem is central to understanding physics of any quantum mechanical system. The eigenvalues and eigenfunctions provides the most detailed information about the system. 

By using a specific basis in the Hilbert space of the quantum system, one can represent the operator Hamiltonian eigenvalue problem as a matrix eigenvalue problem. The process of finding eigenvalues and eigenfunctions reduces to one of matrix diagonalization. Even for a simple system with just one degree of freedom, this matrix is typically infinite dimensional. The sparsity of the Hamiltonian matrix will depend on the nature of both, the terms in the Hamiltonian and the chosen basis.

There is a class of quantum mechanical problems wherein a natural length scale emerges in terms of the fundamental coupling constants appearing in the Hamiltonian. Physics of such problems is typically determined by fluctuations at this length scale. Fluctuations at finer length scales helps improve the accuracy of physical results. There is a second class of quantum mechanical problems wherein coupling between fluctuations at multiple length scales influences the physics of the problem. Relativistic quantum field theory problems fall into this class. A  renormalization group based approach needs to be adopted for dealing with problems involving multiple length scales. In this paper, we focus on the first class of problems.

We will use Daubechies wavelets to quantify the notion of fluctuations at varying length scales. Each element in the basis is labelled by its location and resolution and has compact support. The different terms in the Hamiltonian matrix generated using this basis represent couplings between fluctuations of different length scales. The resulting matrix will allow a natural volume and resolution truncation. An approximate estimate of the eigenvalues will be made using the truncated Hamiltonian matrix. 

The paper is organized as follows. In Section 2, we introduce the Daubechies wavelet basis and establish its key properties essential for analyzing quantum mechanical problems. In Section 3, we illustrate this approach using the simple harmonic oscillator. Section 4 concludes with the summary of key results.

\section{Construction of Wavelet Basis}\label{II}
%%%%%%%%%%%%%%%%%%%%%%%%%%%%%%%%%%%%%%%%%%%%%%%%%%%
Daubechies wavelets arose out the work  \cite{dablec} of the Dutch mathematician, Ingrid Daubechies. They provide a orthogonal and complete basis in the space of square integrable functions. Unlike other types of wavelet basis, Daubechies wavelet basis have finite support and limited smoothness. 

In this section we summarize the key results in theory of Daubechies wavelet required for analysing quantum mechanical problems. We base this section on the work of Wayne Polyzou \cite{polyzoumainart}\cite{polyzoumostused} whose work has been exploring the use of Daubechies Wavelet in the context of relativistic quantum field theory problems.

We begin by defining the translation operator $T$ and scale transformation operator $D$ in the space of square integrable functions. $T$ translates the function , say $f(x)$, by unit length towards right.

\begin{equation}    \label{T}
    Tf(x) = f(x-1)
\end{equation}

The scaling operator, $D$ , squeezes the function it acts on by a factor of $2$, in such a way that the norm ($\int |f(x)|^2 dx$), remains the same. The operator $D$, therefore, changes the scale of the function by decreasing its support size by a factor of $2$. 
\begin{equation}    \label{D}
    Df(x) = \sqrt{2}f(2x)
\end{equation}

These operators can be visualized by the following diagram : \newline

\begin{figure}[H]
    \centering
    \includegraphics[scale=0.4]{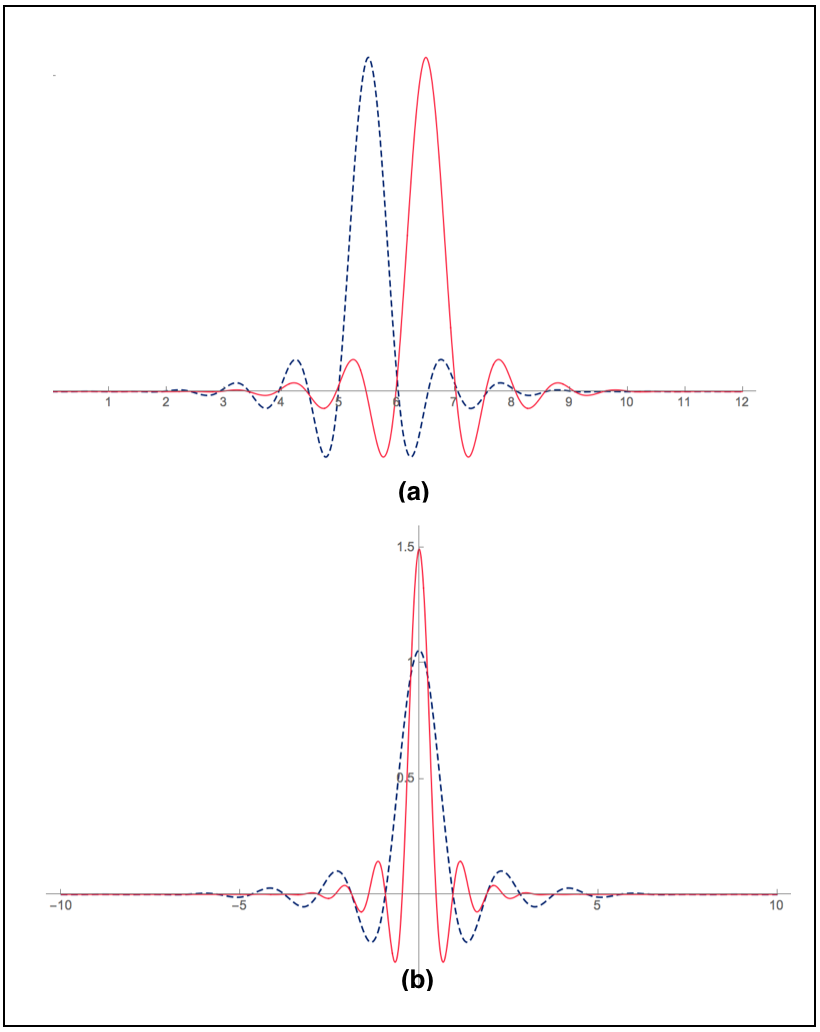}
    \caption{{Comparison of the function (dotted lines indicate the initial function) when acted by (a) translation operator, and (b) dyadic scale operator(final function indicated in solid lines)}}
    \label{fig:my_label}
\end{figure}
 
\begin{figure}[H]
    \centering
    \includegraphics[scale=0.4]{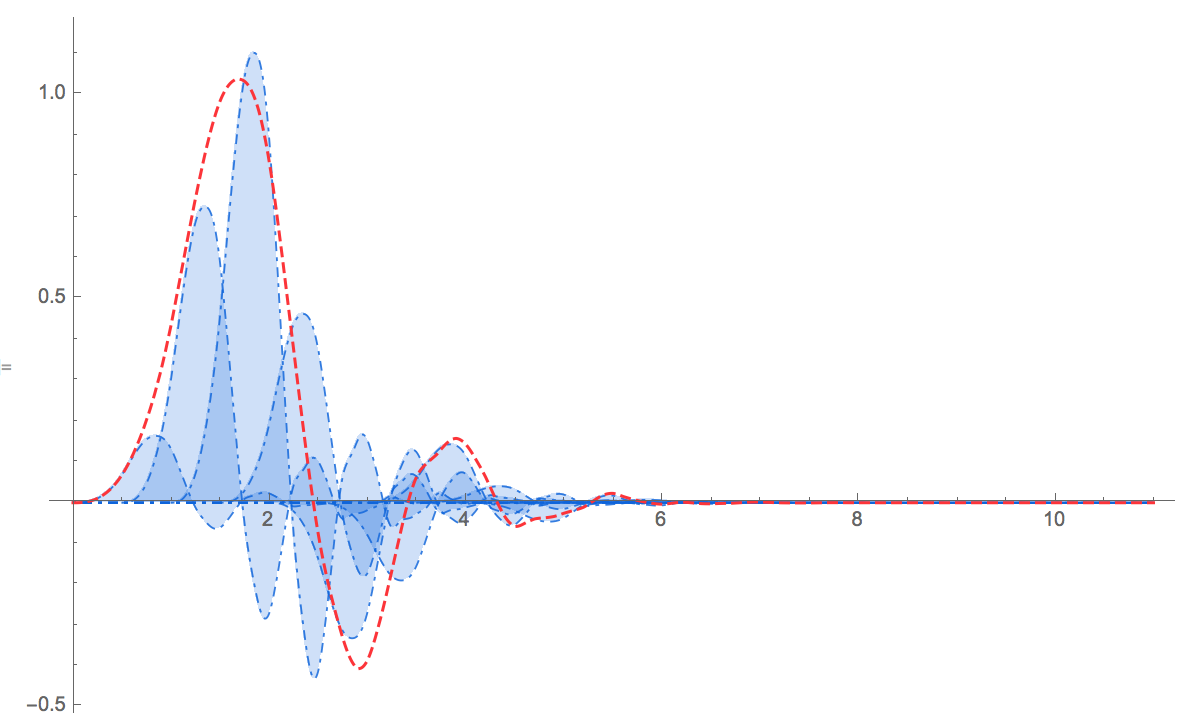}
    \caption{Visual representation of renormalization equation for $K=6$. The red-dotted line is the scaling function which is a linear combination of $12$ translated, half-scaled copies of itself}
    \label{fig:RG}
\end{figure}

Next, one defines the mother scaling function, as a linear combination of $2K$ translated and scaled copies of itself, through the so called 'renormalization' equation

\begin{equation}    \label{renormalization eqn}
    s(x) = \sum_{i=0}^{i=2K-1} h_i D T^i s(x)
\end{equation}
such that $s(x)$ is normalized to one.
\begin{equation}    \label{normalization eqn}
    \int s(x) dx = 1
\end{equation}
Thus to define the scaling function, a choice in the value of $K$ has to be made. This is visually represented in fig.\ref{fig:RG}.  

For a given $K$, the coefficients $h_i$, in eq.\ref{renormalization eqn} are unique. These coefficients are known as filter coefficients, and are solutions of system of equations \ref{hk condn 1},\ref{hk condition 2}, which are a consequence of renormalization equation and normalization condition); and orthonormal conditions(eq.\ref{oss}-\ref{oww})

\begin{equation}        \label{hk condn 1}
    \sum_{i=0}^{2K-1} h_i = \sqrt{2}
\end{equation}

\begin{equation}        \label{hk condition 2}
    \sum_{i=0}^{2K-1}h_{i}h_{i-2j} = \delta_{j0}
\end{equation}

\begin{table}[H]
    \centering
    \begin{tabular}{|c|c|c|c|}
        \hline
        $h_n$ & $K = 1$ & $K = 2$ & $K = 3$   \\
        \hline
        $h_0$ & $1/\sqrt{2}$ & $(1+\sqrt{3})/4\sqrt{2}$ & $(1+\sqrt{10}+\sqrt{5+2\sqrt{10}})/16\sqrt{2}$   \\
        \hline
        $h_1$ & $1/\sqrt{2}$  & $(3+\sqrt{3})/4\sqrt{2}$ & $(5+\sqrt{10}+3\sqrt{5+2\sqrt{10}})/16\sqrt{2}$    \\
        \hline
        $h_2$ & $0$ & $(3 - \sqrt{3})/4\sqrt{2}$ & $(10-2\sqrt{10}+2\sqrt{5+2\sqrt{10}})/16\sqrt{2}$    \\
        \hline
        $h_3$ & $0$ & $(1-\sqrt{3})/4\sqrt{2}$ & $(10-2\sqrt{10}-2\sqrt{5+2\sqrt{10}})/16\sqrt{2}$    \\
        \hline
        $h_4$ & $0$ & $0$ & $(5+\sqrt{10}-3\sqrt{5+2\sqrt{10}})/16\sqrt{2}$   \\
        \hline
        $h_5$ & $0$ & $0$ & $(1+\sqrt{10}-\sqrt{5+2\sqrt{10}})/16\sqrt{2}$    \\
        \hline
    \end{tabular}
    \caption{Filter Coefficients for Daubechies Wavelet for different $K$ values}
    \label{tab:my_label}
\end{table}

The scaling function has compact support, i.e., it is non-zero on a interval of finite length. For $s(x)$, the size of the support is $2K-1$, and the domain where $s(x)$ is non-zero is ($0,2K-1$).Thus the support increases with increasing value of $K$.

The Daubechies function is a non-polynomial function, and the value of the function at the individual points is obtained by making use of eq.\ref{renormalization eqn}, and the compact support of wavelet function\cite{doi:10.1063/1.168556}.The analyticity of the wavelet function depends on $K$. For $K=2$, the Daubechies function is continuous but not differentiable, for $K=4$, the Daubechies function is singly differentiable, for $K=6$, the function is doubly differentiable,and so on.

\begin{figure}[H]
    \centering
    \includegraphics[scale=0.45]{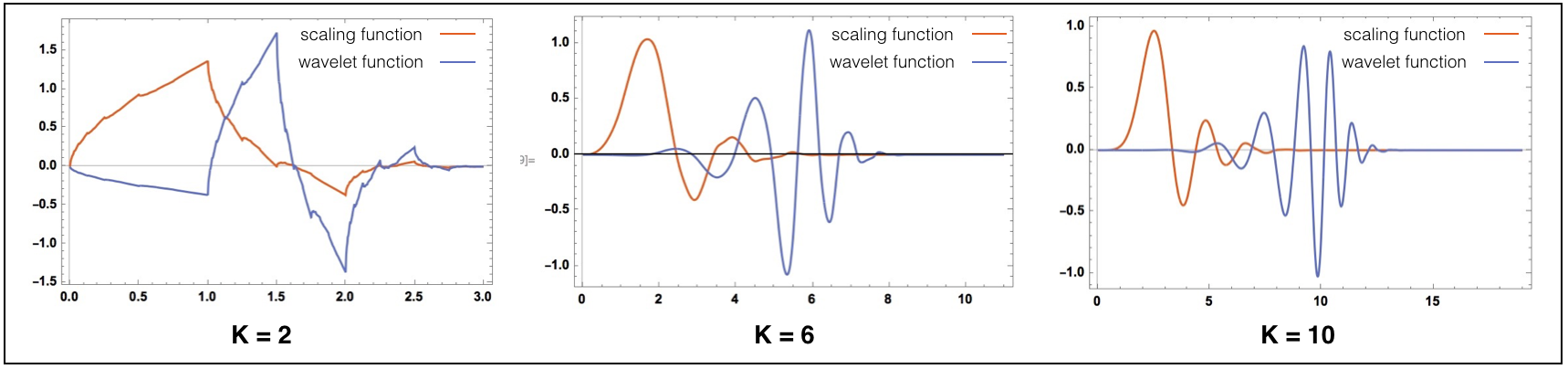}
    \caption{Daubechies Wavelet function for different values of $K$}
    \label{fig:dbK}
\end{figure}

From the mother scaling function $s(x)$, one constructs a basis element with a different scale. We construct the space $\mathscr{H}_k$ of resolution $k$ by 

\begin{equation}    \label{Hk}
    \mathscr{H}^k = \{f(x)| f(x)  = \sum_{-\infty}^{\infty} c_n s^k_n(x) , |c_n^2|<\infty \}
\end{equation}

where

\begin{equation}        \label{skn}
    s^{k}_{n}(x) := D^k T^n s(x)
\end{equation}

The basis elements are labeled by the translation (location) parameter $n$ and scaling (resolution) parameter $k$. In comparision to the mother scaling function, their support is smaller by factor $2^k$. 
\newline

\[s^k_n(x) \ne 0  \quad  \quad \forall x \in     \Big{(} \frac{(0 - n)}{2^k} ,\frac{ (2K-1 - n)}{2^k} \Big{)}\]
\newline
\[\Rightarrow \mbox{Support size}  = \frac{2K-1}{2^k}  \]
 
From the above definition it can be shown that 

\[ \mathscr{H}^{k+1} \subset \mathscr{H}^{k} \]
In general,
\[ \mathscr{H}^{k+m} \subset \mathscr{H}^{k}\]
for $m>0$.
As a next step, one defines the mother wavelet function:

\begin{equation}        \label{mother wavelet}
    w(x) = \sum_{i=0}^{2K-1} g_i D T^i s(x)
\end{equation}

where 
\begin{equation}        \label{mw coeff}
    g_i = (-1)^i h_{2K-1-i}
\end{equation}

Similar to the space $\mathscr{H}^{k}$ ,  the space $\mathscr{W}^{k}$ of resolution $k$ is constructed: 

\begin{equation}        \label{Wk}
    \mathscr{W}^k = \{f(x)| f(x)  = \sum_{-\infty}^{\infty} c_n w^k_n(x) , |c_n^2|<\infty \}
\end{equation}where
\begin{equation}        \label{wkn}
    w^{k}_n(x) := D^k T^n w(x)
\end{equation}The mother wavelet function is defined in a way to ensure that $\mathscr{W}^k$ is the orthogonal complement of $\mathscr{H}^{k}$ in $\mathscr{H}^{k+1}$ i.e. \newline
\begin{equation}        \label{H_{k+1}}
 \mathscr{H}^{k+1}= \mathscr{H}^k \oplus \mathscr{W}^k    
\end{equation}
Eq. \ref{H_{k+1}} can be used recursively to show that 
\begin{equation}    \label{L2R}
    L^2(\mathbb{R}) = \mathscr{H}^{k} \oplus \mathscr{W}^{k+1} \oplus \mathscr{W}^{k+2} \oplus \mathscr{W}^{k+3} \oplus ...  
\end{equation}
where $L^2(\mathbb{R})$ represents Hilbert space of square integrable functions. This can be summarised by the following Euler diagram
\begin{figure}[H]
    \centering
    \includegraphics[scale=0.40]{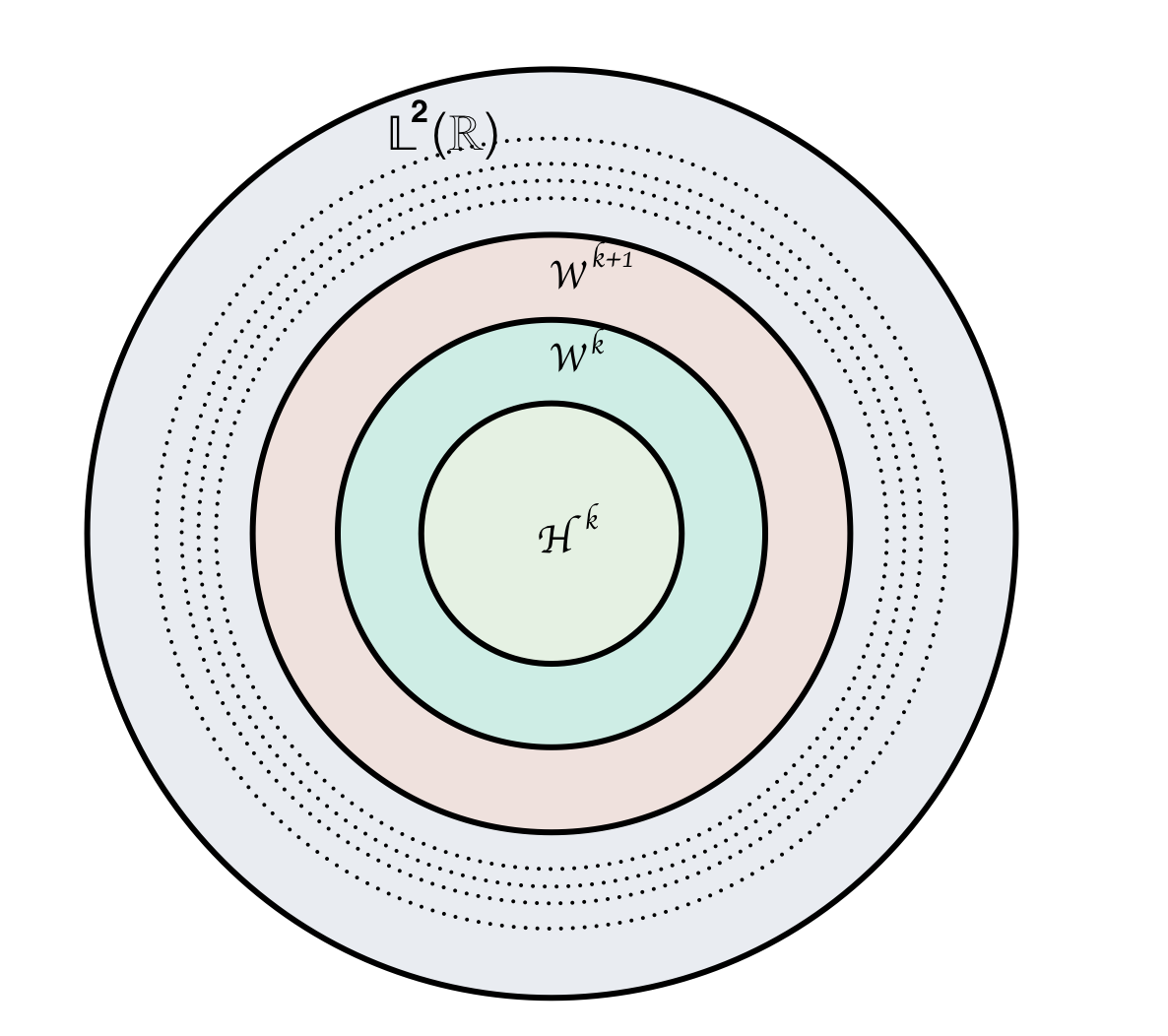}
    \caption{Spanning of Hilbert space by wavelet basis. In this Euler diagram the Hilbert space is spanned by the circular region- $\mathscr{H}^{k}$,  the annular region - $\mathscr{W}^{k+1}$ ,$\mathscr{W}^{k+2}$ and so on}
    \label{fig:l2r}
\end{figure}

From eq.\ref{L2R}, it follows that the basis elements that will span $L^2(\mathbb{R})$ will be :
\begin{equation}
    \{s^k_n\}^{\infty}_{n=-\infty} \cup \{w^m_n\}^{\infty,\infty}_{n=-\infty,m=k}
\end{equation}

where the basis elements satisfy the following orthonormal conditions:
\begin{equation}    \label{oss}
 \int s^k_m(x) s^k_n(x) dx = \delta_{mn}   
\end{equation}      \label{ss}
 
\begin{equation}    \
    \int s^k_n(x) w^{k+l}_{m}(x) dx = 0  , l\geq 0
\end{equation}      \label{osw}

\begin{equation}        \label{oww}
    \int  w^k_n(x) w^l_m(x) dx = \delta_{mn} \delta_{kl}
\end{equation}

\section{Simple Harmonic Oscillator}\label{III}
%%%%%%%%%%%%%%%%%%%%%%%%%%%%%%
We now illustrate the approach to analyzing quantum mechanical problem using simple harmonic oscillator as an example. While this problem has been studied before using Daubechies wavelets\cite{hamosc}, the approach presented here has carries a perspective which allows the reader to begin applying this technique to other problems in quantum mechanics.

The Schrodinger equation for simple harmonic oscillator is given by
\begin{equation}    \label{hamo}
    \frac{-\hbar^2}{2m} \frac{d^2 \psi (x)}{dx^2} + \frac{1}{2} kx^2\psi (x) = E \psi(x)
\end{equation}
The fundamental constant in the problem generate a natural length scale of $(\frac{{\hbar}^2}{mk})^{1/4}$
There is a natural length scale associated with this problem which is due to the presence of the physical constants - $\hbar$, $m$ , $k$ in the above equation. By measuring length in unit of this length scale equation \ref{hamo} is rewritten as :
\begin{equation}    \label{hamo 2}
    \frac{-1}{2} \frac{d^2 \psi (q)}{dq^2} + \frac{1}{2} q^2\psi (q) = \frac{E}{\hbar \omega} \psi(x),\quad \quad \omega = \sqrt{\frac{k}{m}}
\end{equation}
under change of variables $q = (\frac{mk}{\hbar^2})^{1/4} x$. 

To obtain the Hamiltonian matrix, the operator is sandwiched between wavelet basis elements basis,$\{s^k_n\}^{\infty}_{n=-\infty} \cup \{w^m_n\}^{\infty,\infty}_{n=-\infty,m=k}$ , constructed in section $2$ . The resulting Hamiltonian so obtained is of the form
\begin{equation}    \label{rhr}
\small{H_{r' r} = <r'|\hat{H}|r> = \int ( \frac{-r'(q)}{2} \frac{d^2 r'(q)}{dq^2} +\frac{r'(q)}{2} q^2 r(q)) dq}
\end{equation}
where $r(q),r'(q) \in \{s^k_n\}^{\infty}_{n=-\infty} \cup \{w^m_n\}^{\infty,\infty}_{n=-\infty,m=k}$  \newline

This matrix has infinite rows and columns. To diagonalize such matrix, the wavelet basis needs to be truncated to make the Hamiltonian matrix finite. Because there is already a natural length scale associated with the problem, and the form of potential in this length scale is known, one can make a reasonable and an informed choice for the truncation parameters. The choice of the truncation parameters is discussed later in the paper. \newline 

The truncated Hamiltonian matrix so obtained will be:
\begin{equation}
    \{s^k_n\}^{N_s}_{n=-N_s} \cup \{w^m_n\}^{N_w,k+M}_{n=-N_w,m=k}
\end{equation}
where $N_s$, $N_w$, $M$ are the truncation parameters. \newline

\begin{figure}[H]
    \centering
    \includegraphics[scale=0.50]{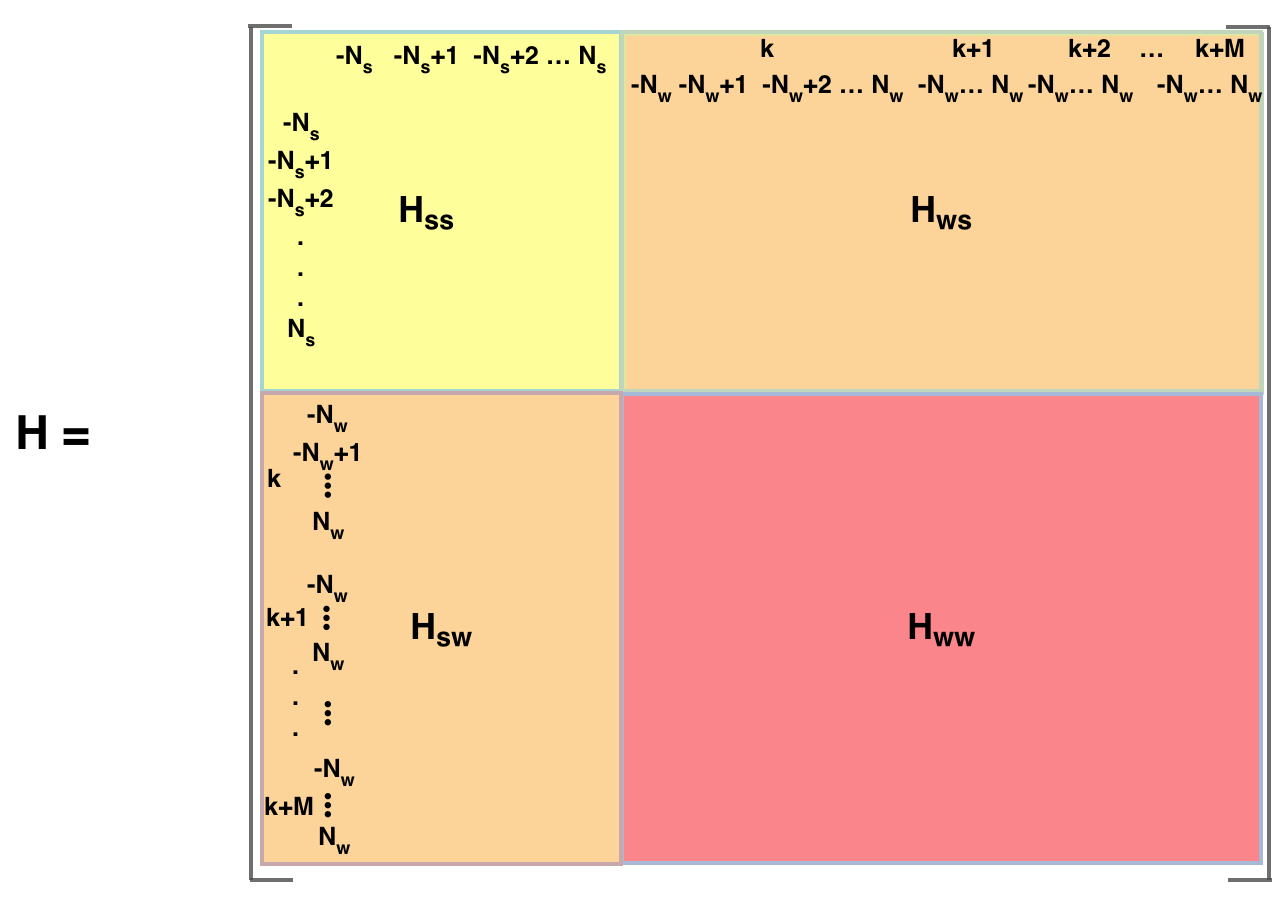}
    \caption{Hamiltonian comprising of various sections}
    \label{fig:hpart}
\end{figure}

This finite Hamiltonian is diagonalized to get the  eigenvalues and the respective eigenvectors. For a better understanding of the matrix, the Hamiltonian matrix can be viewed as comprising of four blocks namely $H_{ss}$,$H_{sw}$,$H_{ws}$, and $H_{ww}$, illustrated in figure \ref{fig:hpart} where the rows and columns are be labelled by the truncation parameters. 

Each of the entry in the truncated hamiltonian is given by the overlap integral (eq. \ref{rhr}) which is either evaluated numerically or by making use of various properties of Daubechies wavelet. In this paper, the overlap integral is calculated using the later approach. A detail analysis of this approach can be found in \cite{polyzoumainart}. The calculation of this overlap integrals, diagonalization of hamiltonian matrix and matrix plots have been carried out using Mathematica. \newline \newline
The value of overlap integrals (eq. \ref{rhr}) decreases as the overlap between $r(q)$ and $r'(q)$ decreases. For example, let $r(q)$ in eq.(\ref{rhr}) be $w^0_{2}$ having a non-zero value in the domain $x\in (2,7)$. For such a basis element the overlap integral decreases as $r'(q)$ moves away from region $(2,7)$ or if the resolution associated with $r'(q)$ as compared with $w^0_{2}$ becomes finer and finer. Since translation and scaling parameters determine the location and resolution of basis elements, they will therefore determine the overlap integral. \newline \newline
 \begin{figure}[H]
    \centering
    \includegraphics[scale=0.35]{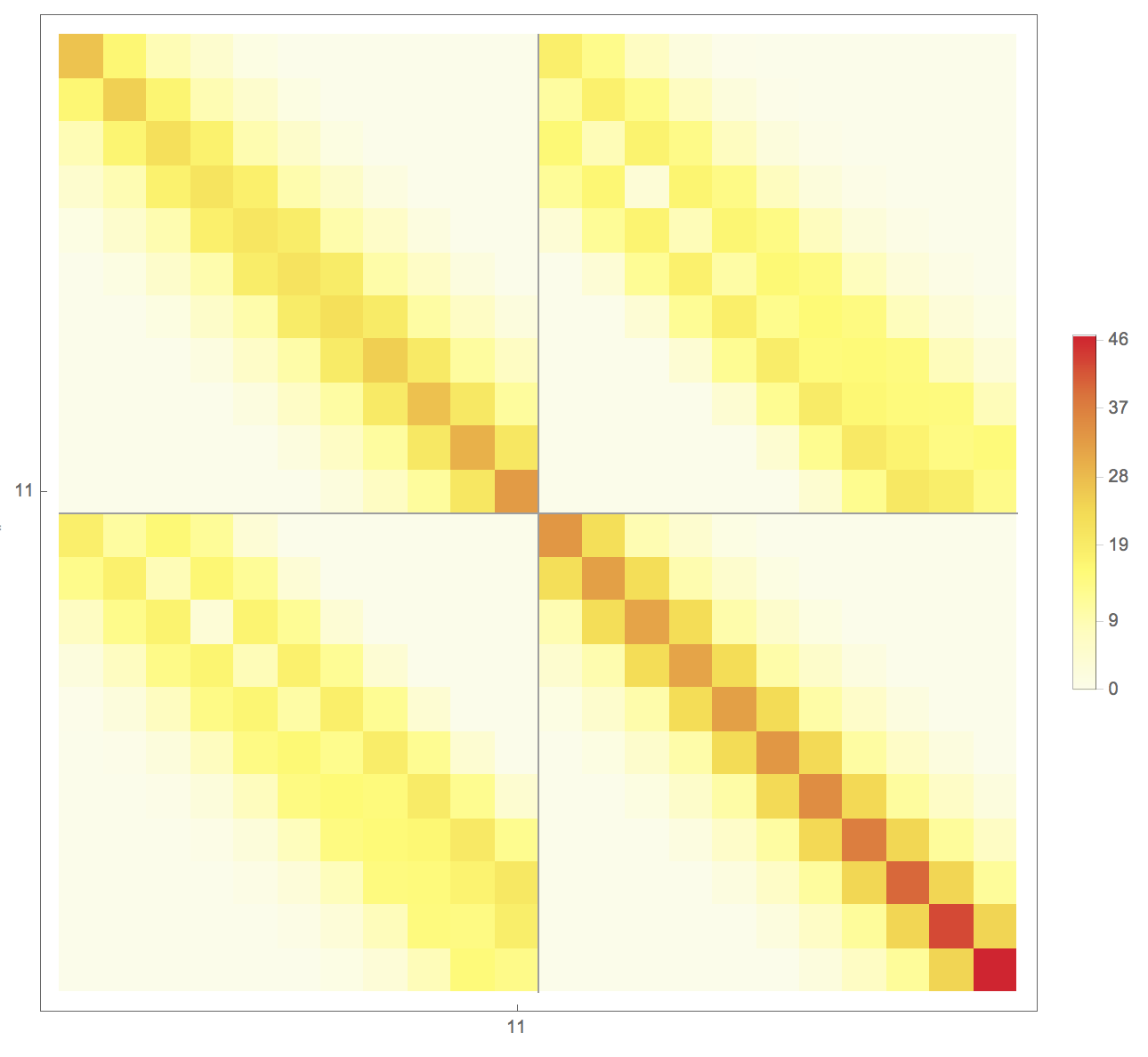}
    \caption{Matrix plot of the Hamiltonian for k=0, M=0. The matrix is purposefully compartmentalized to highlight different compartments of the hamiltonian as illustrated in fig.\ref{fig:hpart},}
    \label{fig:M=0}
\end{figure}
As can be seen in figure \ref{fig:hpart}, the matrix is non-zero along the block diagonal parts of  $H_{ss}$,$H_{sw}$,$H_{ws}$, and $H_{ww}$. This is because in this region, $r(q)$ and $r'(q)$ have maximum overlapping, because they share domain where the basis elements are non-zero. As the translation parameter changes, i.e. as one moves away from the diagonal of these compartments, the overlap between $r(q)$ and $r'(q)$ decreases and the overlap integral decreases.
%%%%%%%%%%%%%%%%%%%%%%%%%%%%%%

\section{Results}\label{IV}
%%%%%%%%%%%%%%%%%%%%%%%%%%%%%%
To calculate the eigenvalues, a choice in the truncation parameters and value of $K$ has to be made. For the truncation parameters, the choice is made for the range of translation - $N_s$and $N_w$, and scaling parameters $k$ and $M$. The optimal choice is obtained the following way- keeping the scaling parameters fixed, say $k=0$ and $M=1$, and choosing $K=3$,  eigenvalues are calculated for varying translation parameters. Choosing $K=3$ and $k=0$ ensures that the basis element corresponding to coarsest scale and the scale of the potential are of the same order.  Having obtained the maximal range for the translation parameters, the eigenvalues are calculated for varying $M$, to see how the eigenvalues change on adding finer basis elements to the truncated basis.

\begin{figure}[H]
    \centering
    \includegraphics[scale=0.50]{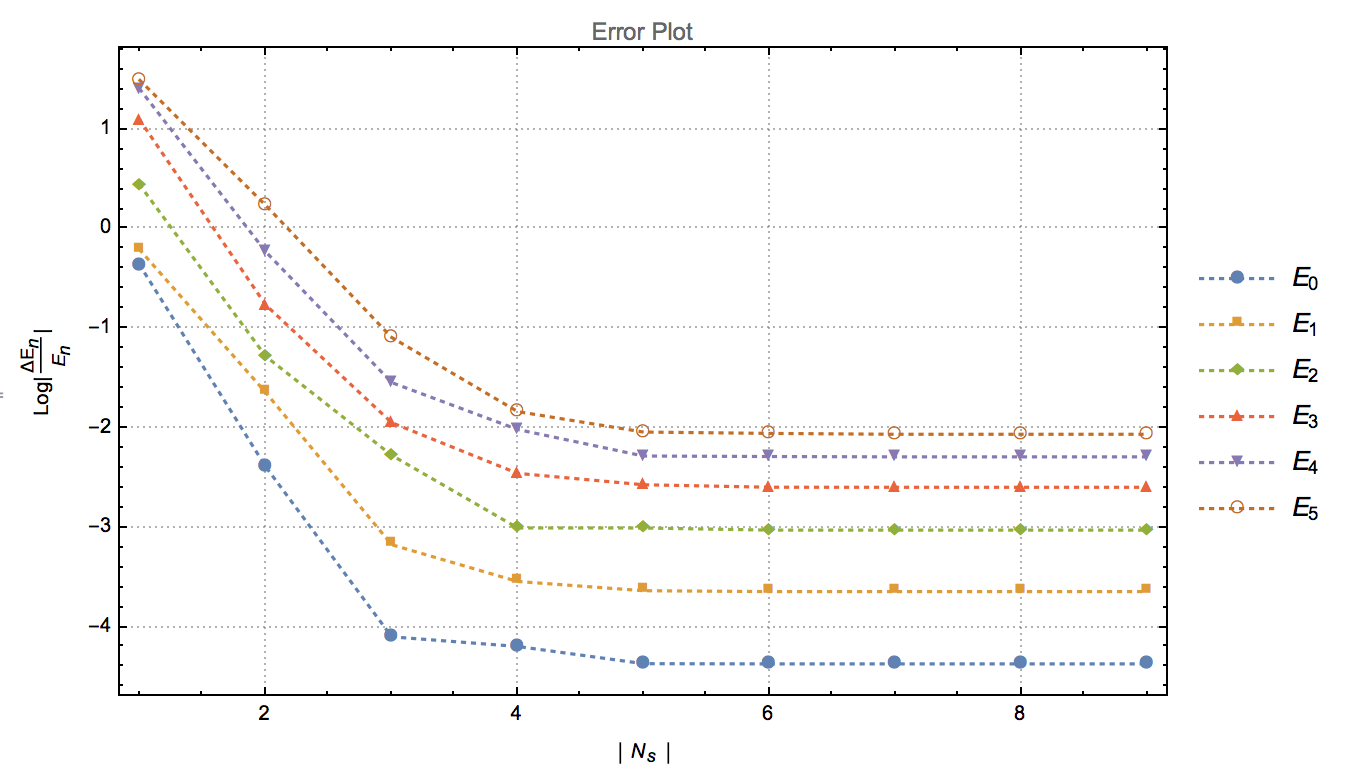}
    \caption{The natural log of relative error versus increasing $N_s$ and $N_w$ value such that $N_s$=$N_w$}
    \label{fig:errplot}
\end{figure}

To obtain a maximal translation range, a choice is made for scaling parameter, $k=0$ and $M=1$, and eigenvalues are calculated for symmetrically varying $N_s$ and $N_w$ values (i.e $|N_s|$=$|N_w|$). The logarithmic error of the calculated eigenvalues is plotted against these parameters then.(see fig. \ref{fig:errplot}). The logarithmic error, log( $|E_{calculated}$-$E_{expected}|$), makes use of expected eigenvalue known from the solution of the simple harmonic oscillator. From the figure \ref{fig:errplot}, it can be seen that the error saturates around a value of $|N_s|$=5. Therefore, $|N_s|$=5, is the optimal translation range, beyond which addition of basis elements do not improve the accuracy of eigenvalues. The saturation of the eigenvalue points out that the fluctuations beyond certain volume play no role in the physics of simple harmonic oscillator.

With the range of truncation parameter fixed to optimal value, $|N_s|$=$|N_w|$=5, eigenvalues are calculated for varying $M$. As $M$ increases, finer resolution basis elements are added to the truncated basis.  An increase in the accuracy of eigenvalue is expected because finer fluctuations of the Hamiltonian are captured with the new truncated basis. The results of calculation of different eigenvalues  for increasing $M$ values is tabulated in table \ref{tab:evalvsM}. 

\begin{table}[H]
    \centering
    \begin{tabular}{ccccc}
        \hline
        $n$ & $Expected$ $E_n$ & $M = 0$ & $M = 1$ & $M = 2$   \\
        \hline
        0 & 0.5 & 0.50634 & 0.50050 & 0.50003  \\
        \hline
        1 & 1.5 &  1.53959 & 1.50341 & 1.50023   \\
        \hline
        2 & 2.5 &  2.62392 & 2.51179 & 2.50082   \\
        \hline
        3 & 3.5 &  3.76781 & 3.52860 & 3.50204   \\
        \hline
        4 & 4.5 &  4.95963 & 4.55616 & 4.50413   \\
        \hline
        5 & 5.5 &  6.21470 & 5.59639 & 5.50740   \\
        \hline
        6 & 6.5 &  7.58448 & 6.65060 & 6.51230   \\
        \hline
        7 & 7.5 &  8.99486 & 7.71812 & 7.51857   \\
        \hline
        8 & 8.5 &  10.2995 & 8.80116 & 8.52570   \\
        \hline
    \end{tabular}
    \caption{Calculated Eigenvalues as $M$ increases.}
    \label{tab:evalvsM}
\end{table}

The table shows an increase in the accuracy of the eigenvalue for increasing $M$. It is also noticeable that with every unit increase in the $M$ value, the accuracy increases more for lower eigenvalues. This is consistent with the earlier results ( see fig.\ref{fig:errplot} ) where lower eigenvalues have a lower saturation value for error. The optimal range affects the accuracy unevenly, therefore in the table smaller eigenvalues have a better precision than larger eigenvalues for the same range of translation parameters. \newline
%%%%%%%%%%%%%%%%%%%%%%%%%%%%%%

\section{Conclusion}\label{V}
%%%%%%%%%%%%%%%%%%%%%%%%%%%%%%
Many problems in quantum mechanics are defined in position space. Examples include simple harmonic oscillator and the Hydrogen atom problem (examples of exactly soluble problems) as well the complex atomic structure problem. Theories of elementary particles and their interactions are best formulated as quantum field theories with the fundamental variables as quantum fields defined on space-time. So it is natural to talk of quantum fluctuations of at varying length scale and understanding the way in which these contribute to determining physics of the problem.

In this paper we have introduced the reader to a multi-scale resolution approach to analyzing quantum mechanical problems using Daubechies wavelet basis. Each element of this basis represents a quantum fluctuation at specific length scale centred at a particular location. The Hamiltonian matrix in this basis clearly displays the extent of coupling between various length scales. We illustrated the approach using one dimensional simple harmonic oscillator. The analysis shows that the fluctuations which match the underlying length scale of the problem determine the overall size of the eigenvalues. The accuracy of the eigenvalues can be improved by adding fluctuations of finer resolution.

We believe that this technique will be show its true potential in analysis of quantum field theories where there is no underlying natural length scale dictated by the coupling constants in the Hamiltonian. In such problems, the Hamiltonian displays coupling of fluctuations across multiple length scales. This multi-scale resolution approach using Daubechies wavelet basis will provide a natural means to implementing renormalization group based schemes which are essential for extracting physics out of such problems. 
%%%%%%%%%%%%%%%%%%%%%%%%%%%%%%

%%References section
\bibliographystyle{unsrt}
\bibliography{edit_1}

\begin{thebibliography}{1}

\bibitem{dablec}
Ingrid Daubechies.
\newblock {\em Ten Lectures on Wavelets}.
\newblock Society for Industrial and Applied Mathematics, 1992.

\bibitem{polyzoumainart}
Wayne Polyzou and Fatih Bulut.
\newblock {Wavelet Methods in Field Theory}: {Wavelets in Field Theory}.
\newblock {\em Few Body Syst.}, 55:561--566, 2014.

\bibitem{polyzoumostused}
Fatih Bulut and W.~N. Polyzou.
\newblock Wavelets in field theory.
\newblock {\em Physical Review D}, 87(11), Jun 2013.

\bibitem{doi:10.1063/1.168556}
Alistair C.~H. Rowe and Paul~C. Abbott.
\newblock Daubechies wavelets and mathematica.
\newblock {\em Computers in Physics}, 9(6):635--648, 1995.

\bibitem{hamosc}
Jason~P. Modisette, Peter Nordlander, James~L. Kinsey, and Bruce~R. Johnson.
\newblock Wavelet based in eigenvalue problems in quantum mechanics.
\newblock {\em Chemical Physics Letters}, 250(5):485 -- 494, 1996.

\end{thebibliography}

%%%%%%%%%%%%%%%%%%%%%%%%%%%%%%

\end{document}